# DATA ANALYSIS ON CREDIT CARD DEBT: RATE OF CONSUMPTION AND IMPACT ON INDIVIDUALS AND THE US ECONOMY

Mayowa Akinwande[1], Alexander Lopez[2], Tobi Yusuf[3], Austine Unuriode[4],Babatunde Yusuf[5], Toyyibat Yussuph[6] and Stanley Okoro[7]

## ABSTRACT

*This paper provides a comprehensive examination of the evolution of credit cards in the United States, tracing their historical development, causes, consequences, and impact on both individuals and the economy. It delves into the transformation of credit cards from specialized merchant cards to ubiquitous financial tools, driven by legal changes like the Marquette decision. Credit card debt has emerged as a significant financial challenge for many Americans due to economic factors, consumerism, high healthcare costs, and financial illiteracy. The consequences of this debt on individuals are extensive, affecting their financial well-being, credit scores, savings, and even their physical and mental health. On a larger scale, credit cards stimulate consumer spending, drive e-commerce growth, and generate revenue for financial institutions, but they can also contribute to economic instability if not managed responsibly.*

*The paper emphasizes various strategies to prevent and manage credit card debt, including financial education, budgeting, responsible credit card uses, and professional counselling. Empirical studies support the relationship between credit card debt and factors such as financial literacy and consumer behavior. Regression analysis reveals that personal consumption and GDP positively impacts credit card debt indicating that responsible management is essential. The paper offers comprehensive recommendations for addressing credit card debt challenges and maximizing the benefits of credit card usage, encompassing financial education, policy reforms, and public awareness campaigns. These recommendations aim to transform credit cards into tools that empower individuals financially and contribute to economic stability, rather than sources of financial stress.*

## KEYWORDS

*Debt, Financial literacy, financial well-being, Economic stability, Credit cards*

## 1. INTRODUCTION

Until the 1950s, credit cards were confined to specific merchants and lacked prominence as a household credit tool. However, the landscape changed when Diners Club introduced the first general-purpose charge card in the early 1950s [1]. This innovation inspired Bank of America to launch BankAmericard, the inaugural general-purpose credit card. By the 1970s, over 100 million BankAmericard-style cards circulated, and Bank of America extended licensing to other banks, ultimately leading to the formation of Visa as a separate entity (Charles & Joanna, 2008).

While credit card payment technology advanced significantly, the practice of borrowing via credit cards didn't gain momentum until the 1980s. High inflation and usury legislation that enforced interest rate caps combined to cause this delayed acceptance. These conditions posed profitability challenges for credit card lending during the 1970s, with many lenders teetering on the brink due to soaring inflation. A pivotal moment arrived in 1978 with the U.S. Supreme Court's ruling in Marquette National Bank of Minneapolis v. First of Omaha Service Corporation. The decision allowed banks to charge interest rates based on their chartering state, even if these rates exceeded usury limits in the states where they offered their services [2]. Essentially, this ruling permitted banks to "export" their interest rate caps, sparking competition among states to attract bank headquarters. South Dakota and Delaware were among the first states to raise their usury limits on interest rates, drawing in credit card issuers (Lukasz, 2021).

Due to this legal change, credit card debt has steadily increased, replacing non-revolving consumer credit, and making the United States a credit card debtor country. The 1990s saw a surge in credit card lending among lower-income and riskier households, causing credit card debt relative to median household income to double every decade until the 2008 financial crisis. By early 2008, households with at least one card had accrued over 20 percent of their income in credit card debt.

This transformation was particularly profound for median-income households, as most of the income growth was concentrated among the top 1 percent of earners who used credit cards less frequently. As a result, credit card lending became a vital financial tool for most American households. The pivotal moment in the credit card lending boom was the relaxation of usury laws, driven by the Marquette decision, which allowed banks to set interest rates based on their chartering state. This regulatory change, combined with increased credit card usage among lower-income households, led to a significant upsurge in credit card debt, establishing it as a critical component of the financial landscape for most U.S. households (Lukasz, 2021).

Credit cards have emerged as a primary means of conducting and funding transactions within the United States. (Tufan & Lucia, 2006). Credit cards offer several advantages, serving as a convenient payment method, a valuable tool for learning financial responsibility, an emergency resource, a means to establish a positive credit history, and a way to enhance future credit opportunities [3]. However, if credit cards are mishandled or misused, the drawbacks can lead to significant financial repercussions (Antony, 2018).

The convenience of credit cards may entice users to exceed their financial means. Accumulating excessive credit card debt and missing payments can harm an individual's credit score, making it challenging to access credit in the future. Furthermore, individuals who lack financial experience might not fully grasp the compounding impact of interest rates on their debt. Insufficient knowledge about credit and personal finances can expose some users to higher financial risks, potentially resulting in substantial and unmanageable debt burdens, ultimately leading them into financial distress (Antony, 2018).

Traditional economic rationales for credit card debt revolve around assessing the comparative expense of borrowing through credit cards as opposed to other sources of credit. While utilizing credit cards for borrowing may seem costly on the surface, some economists argue that other alternatives could potentially be even more financially burdensome. The cost of borrowing is not limited solely to the interest rate but also encompasses factors like the difficulty of securing alternative financing and the expenses associated with transitioning between various credit agreements (Charles & Joanna, 2008).

In the second quarter of 2023, the total debt held by households increased by $16 billion, reaching a total of $17.06 trillion, as reported in the most recent Quarterly Report on Household Debt and Credit. Notably, credit card balances experienced substantial growth, surging by $45 billion to reach a record high of $1.03 trillion. Additionally, other forms of debt, including retail credit cards, various consumer loans, and auto loans, collectively increased by $15 billion and

$20 billion, respectively [4]. On the other hand, student loan balances decreased by $35 billion, bringing the total to $1.57 trillion, while mortgage balances remained largely stable at $12.01 trillion (Federal Reserve bank of New York, 2023).

Maintaining credit card balances over time appears to be an expensive method of financing one's expenses. This high cost has prompted two general avenues of research aimed at elucidating the underlying reasons behind holding such revolving debts. The first, which follows a more conventional approach, endeavors to elucidate credit card indebtedness through explanations rooted in costs. The second path, characterized by its behavioral focus, seeks to account for credit card debt through psychological factors, such as issues related to self-control (Charles & Joanna, 2008).

Credit cards rank among the most prevalent and extensively utilized financial tools in the United States, with over 175 million Americans possessing at least one credit card. Throughout the COVID-19 pandemic [5], credit cards assumed a critical role, serving both as a means of accessing credit during emergencies and as a preferred payment method as online transactions surged (Bureau of Consumer Financial Protection, 2021). This paper investigates the connection between credit card debt and household consumption in United States. Furthermore, it assesses the influence of credit card debt on the U.S. economy.

This paper aligns with established studies on credit card dynamics, emphasizing the substantial impact of credit card debt on both individual well-being and the broader economy. Consistent themes include the significance of financial literacy, the role of credit cards in economic crises, and the diverse consequences of credit card debt. While some nuanced differences exist, the collective body of research contributes to a comprehensive understanding of the intricate dynamics surrounding credit card usage and its implications.

## 2. LITERATURE REVIEW

### 2.1. Causes of Credit Card Debt

Credit cards have become an integral part of the financial landscape in the United States, offering convenience, flexibility, and purchasing power to millions of consumers. While they offer numerous advantages, credit cards can also lead to significant financial challenges when not managed responsibly. In this article, we will explore the primary causes of credit card debtin the USA, shedding light on the factors that contribute to this pervasive issue.

One of the major contributors to credit card debt in the USA is economic factors, including unemployment, underemployment, and stagnant wages. Economic factors continue to play a substantial role in driving credit card debt in the USA [6]. Economic downturns, such as the global financial crisis in 2008 and the COVID-19 pandemic, can lead to job losses, reduced income, and financial instability. During such crises, individuals often turn to credit cards to cover essential expenses, contributing to the accumulation of debt (U.S. Bureau of Labor Statistics, 2021).

The high cost of healthcare in the USA remains a significant contributor to credit card debt. Even with insurance, individuals can face substantial out-of-pocket expenses for medical treatments, prescription medications, and unexpected medical emergencies [7]. Medical debt often drives individuals to rely on credit cards to bridge the gap (Kaiser Family Foundation, 2021).

The culture of consumerism in the USA often encourages overspending. The consumer culture in the USA promotes overspending as individuals strive to keep up with societal expectations. Impulse buying, lavish lifestyles, and a desire to "keep up with the Joneses" can lead to excessive credit card use. This culture of consumerism continues to encourage people to spend beyond their means, resulting in credit card debt [8]. Many individuals succumb to the allure of purchasing beyond their means, leading to credit card debt. Impulse buying, lavish lifestyles, and a desire to keep up with others can exacerbate this issue (Liao et al, 2020).

Financial illiteracy remains a pervasive issue, with many Americans lacking basic financial knowledge. Understanding credit card interest rates, minimum payments, and the consequence of carrying a balance is crucial for responsible credit card use [9]. The absence of financial education can lead to poor financial decisions and increased credit card debt (Lusardi & Mitchell, 2014).

Credit card companies often require only minimum payments on outstanding balances, which can create a false sense of affordability. However, high-interest rates on unpaid balances can lead to long-term debt accumulation and substantial interest costs. These high-interest rates can make it challenging for individuals to pay down their debt effectively (Consumer Financial Protection Bureau, 2021).

Unforeseen emergencies, such as car repairs, home maintenance, or sudden job loss, can leave individuals with no choice but to rely on credit cards to cover these unexpected expenses. Without an emergency fund, individuals may accumulate credit card debt when faced with unexpected financial challenges (PRC, 2020).

## 2.2. Impact of Credit Card Debt on Individual

Consumers encounter economic motives encouraging them to possess and utilize credit cards, such as the allure of rewards programs or the benefit of an interest-free grace period. Asignificant body of research examining consumer choices in the credit card industry emphasizes these incentives and underscores their substantial role in predicting consumer behavior (Lam and Ossolinski, 2015). Anxiety and concern about money are frequently caused by credit card debt. The constant worry about high balances, interest rates, and monthly payments can take a toll on an individual's mental and emotional well-being. A study by Northwestern Mutual found that financial anxiety affects 73% of Americans, with credit card debt as a leading source of stress.

High credit card debt can hinder individuals' ability to save money and plan for retirement. Funds that could have been directed towards savings or investments are often diverted to cover credit card payments and interest charges. The lack of savings can lead to financial vulnerability in emergencies and delayed retirement.

Credit card debt can negatively impact an individual's credit score, making it harder to access favorable financial opportunities. A lower credit score can result in higher interest rates on loans, difficulty securing mortgages or rental agreements, and limited access to credit when needed (Consumer Financial Protection Bureau, 2021).

Financial strain due to credit card debt can put a significant strain on personal relationships, including marriages and family dynamics. Disagreements over money and debt management are common and can lead to marital conflicts and even divorce.

The stress associated with credit card debt can have physical health consequences, including sleep disturbances, headaches, and even heart problems. A study published in JAMA Network Open found a link between credit card debt and an increased risk of psychological distress (Niedzwiedz, C. L., et al. (2019). Individuals burdened by credit card debt often experience a reduced quality of life. They may be unable to afford vacations, pursue hobbies, or invest in personal development due to their financial obligations.

## 2.3. Impact of Credit Card on the Economy

Credit cards have become an integral part of American financial life, revolutionizing the way people make transactions and manage their finances. The widespread use of credit cards has not only reshaped consumer behavior but has also had a substantial impact on the broader US economy. In this article, we will delve into the multifaceted effects of credit cards on the US economy, examining both the positive and negative aspects, supported by citations and references.

Credit cards have undergone a remarkable transformation since their inception. Initially, they were limited to specific merchants and primarily aimed at enhancing convenience during in- store purchases. However, over time, credit cards evolved into versatile financial instruments that enable consumers to make a wide range of transactions, both online and offline.

One of the most significant impacts of credit cards on the US economy is their role in stimulating consumer spending. Credit cards provide consumers with convenient access to credit, allowing them to make purchases even when they lack the necessary cash on hand. This increased spending serves as a driving force behind economic activity, supporting businesses and contributing to overall economic growth. According to a study by the Federal Reserve Bank of New York, households with access to credit cards tend to exhibit higher spending levels compared to those without such access (Federal Reserve Bank of New York, 2017).

Credit cards have played a pivotal role in the expansion of e-commerce. They offer a secure and convenient payment method for online shoppers, enabling businesses to reach a global customer base [10]. This growth of e-commerce has led to increased revenues for online retailers and created job opportunities in the digital economy. Statista estimates that e-commerce sales in the United States will reach $563.4 billion in 2021, underscoring the substantial economic impact of online shopping (Statista, 2021).

Credit card companies and financial institutions generate significant revenue through fees, interest charges, and interchange fees paid by merchants. These revenues contribute to the profitability of financial institutions and support the growth of the financial sector [11]. For instance, JPMorgan Chase, one of the largest credit card issuers in the US, reported credit card revenue of $36.4 billion in 2020, illustrating the financial sector's reliance on credit card-related income (JPMorgan, 2021).

## 2.4. Strategies to Prevent Credit Card Debt

For many Americans, credit card debt is still a major source of financial challenge. Managing credit cards responsibly is essential to avoid falling into debt traps that can lead to financial stress and long-term financial consequences. Below are strategies to prevent credit card debt in the United States:

### 2.4.1. Financial Education and Literacy

Improving financial education and literacy is a fundamental strategy to prevent credit card debt. Individuals need a clear understanding of how credit cards work, including interest rates, fees, and minimum payments. Financial literacy programs and initiatives can equip consumers with the knowledge and skills needed to make informed financial decisions [9].

### 2.4.2. Budgeting and Financial Planning

Creating a budget and adhering to it is crucial for responsible credit card use. Budgets help individuals track their income and expenses, allowing them to allocate funds for essential needs and savings while limiting discretionary spending. Budgeting apps and tools can simplify the process and promote financial discipline [10].

### 2.4.3. Emergency Savings Fund

Establishing an emergency savings fund is a proactive strategy to prevent the need for credit card reliance during unexpected financial crises. A readily accessible fund can cover unforeseen expenses like medical bills or car repairs without resorting to credit cards [7].

### 2.4.4. Responsible Credit Card Use

Responsible credit card use involves paying bills on time, in full, and avoiding unnecessary spending. To prevent credit card debt, individuals should only charge what they can afford to pay off each month. Additionally, setting up automatic payments can help ensure timely bill settlement [13].

### 2.4.5. Reduce Credit Card Usage

Reducing the number of credit cards in possession can be an effective strategy. Fewer credit cards mean fewer opportunities for overspending. Individuals can consider canceling unused or high-fee cards and retaining only those with favorable terms [15].

### 2.4.6. Seek Low-Interest Rate Cards

When choosing credit cards, individuals should prioritize those with lower interest rates. Lower interest rates can significantly reduce the cost of carrying a balance, making it easier to pay off any debt accumulated [18].

### 2.4.7. Regularly Monitor Credit Card Statements

Regularly monitoring credit card statements helps individuals identify any unauthorized or fraudulent charges promptly. It also allows them to track their spending patterns and detect any signs of overspending early [17].

### 2.4.8. Consult a Financial Professional for Counselling

In cases where credit card debt becomes unmanageable, seeking professional financial counseling can be beneficial. Credit counselors can provide guidance on debt repayment strategies, negotiate with creditors, and create tailored financial plans to regain control of finances [13].

## 2.5. How to Manage Credit Card Debt

A comprehensive budget is the foundation of effective debt management. Credit card users can create a comprehensive budget to manage their debts and start by listing all sources of income and categorizing your expenses. Identify areas where you can cut discretionary spending and allocate those funds towards debt repayment [1].

Credit card users should focus on paying off high-interest credit card debts first. These debts accumulate interest quickly, making them costlier in the long run. Consider the snowball or avalanche method for debt repayment, depending on your preference and financial situation [14].

Debt consolidation involves combining multiple debts into a single, lower-interest loan. This simplifies payments and can reduce the overall interest you pay. Credit card users should explore options like balance transfer credit cards or personal loans for consolidation [16].

When struggling with credit card debt, it is necessary to contact creditors to negotiate for lower interest rates, reduced fees, or a more manageable repayment plan. Creditors may be willing to design new plans to recover their money [15].

Debt management programs offered by reputable credit 7ounseling agencies can help individuals to consolidate debts and negotiate with creditors. These programs often result in lower interest rates and a structured repayment plan [7].

While repaying existing debt, it is necessary to avoid accumulating more debt. Cut back on credit card usage, especially for non-essential purchases. Individuals should only charge what they can afford to pay off in full each month to prevent further debt accumulation (Investopedia, 2021).

Having an emergency fund can prevent reliance on credit cards when unexpected expenses arise. Households should aim to save three to six months' worth of living expenses in an easily accessible account (Investopedia, 2021).

To manage credit card debt, households and individuals need to seek professional 7ounseling on finance. Professional financial 7ounseling can provide personalized guidance on managing credit card debt. Non-profit organizations like the National Foundation for Credit Counselling (NFCC) offer free or low-cost 7ounseling services (National Foundation for Credit Counselling. (2021).

Individuals need to regularly check their credit reports for inaccuracies and signs of identity theft. Monitoring your credit can help you stay informed about your financial situation and detect any issues early (Federal Trade Commission, 2021).

## 2.6. Empirical Review

(Wesley, et al., 2012) examined the credit card risk behavior of consumers. The study indicated that college students often demonstrate a lack of proficiency in responsibly utilizing a credit card. Additionally, the study unveiled that having a greater number of credit cards heightens the likelihood of engaging in risky financial behaviors. Furthermore, it found that students who claimed to be aware of the interest rates charged by credit card issuers were less inclined to participate in risky financial activities.

Yazgan and Yilmazkuday (2007) examined the effects of credit and debit cards on the currency demand. The research employed GMM estimation to examine how credit and debit cards influence the amount of currency in circulation. In consistency with theoretical expectations, the study discovered that an uptick in credit and debit card usage correlates with a reduction in the demand for physical currency [17]. Furthermore, the impact of debit card usage on currency demand is more pronounced than that of credit cards. Additionally, the study revealed that the influence of credit cards primarily stems from purchases, while debit cards have a more substantial effect through cash withdrawals.

In their 2015 study, Lam and Ossolinski employed a method to assess consumers' willingness to accept a surcharge for card usage, which serves as an indicator of the consumer surplus derived from using their cards. This approach considers both the perceived monetary advantages of card usage, such as rewards points and interest-free periods, as well as the non-monetary benefits, such as convenience and widespread acceptance. The study discovered significant variability among respondents in terms of the value they attributed to using their cards. Generally, individuals with rewards cards exhibited a greater willingness to pay a surcharge, and this inclination remained consistent even after accounting for demographic characteristics and stated preferences.

By integrating the technology acceptance model (TAM) and the theory of perceived risk,(Hoang, et al., 2020) established a theoretical model for consumer behavioral intention, which is evaluated on the planned usage of credit cards in Vietnam. According to the results of structural equation modeling, elements including perceived risk, perceived usefulness, social influence, and perceived ease of use had a substantial impact on consumers' intents to use a credit card. Only perceived danger among these considerations prevented people from using their credit cards as planned. This impression of risk included a range of aspects, such as dangers to one's psychological well-being, finances, performance, privacy, time, relationships, and security.

## 2.7. Theoretical Framework

### 2.7.1. Theory of Planned Behavior

The Theory of Reasoned Action (TRA), which was initially presented by Fishbein and Ajzen in 1975, serves as the foundation for the Theory of Planned Behavior (TPB). TRA posits that behavior can be predicted by considering individuals' attitudes toward a specific behavior and the subjective norms surrounding it. These factors combine to form behavioral intentions that ultimately shape actual behavior. Building upon this framework, Ajzen extended TRA in 1991 to create the TPB. TPB introduces a novel determinant of behavior, marking a departure from TRA.

In addition to attitudes and subjective norms, TPB incorporates the concept of perceived behavioral control (PBC) to play a pivotal role in the formation of behavioral intentions and subsequent actions. PBC addresses a crucial gap in the original TRA model by accounting for personal control over behavior. Researchers suggest that the more positive one's attitude is toward a particular behavior, the more favorable the subjective norm, and the greater the perceived behavioral control, the stronger the resulting behavioral intention (Hrubes, Ajzen, & Daigle, 2001).

Ajzen (2008) has further expanded TPB by emphasizing its adaptability to include additional predictors, albeit with specific criteria. These supplementary predictors must be pertinent to the behavior in question, conceptually distinct from existing TPB variables, and have the potential to causally influence both behavioral intentions and actual behavior. Moreover, it is recommended that any incorporation of new predictors should not only align with theoretical principles but also be supported by empirical evidence and applicable to other subjects explored in social science research.

Considering these criteria, financial literacy is a predictor within the TPB framework. Financial literacy is a clearly defined concept that has undergone extensive examination in various social science disciplines, including psychology, economics, and sociology. It specifically assesses knowledge related to personal finance; a realm distinct from the original variables of the TPB. Furthermore, existing research has established a connection between lower levels of financial literacy and higher levels of credit card debt, implying that financial literacy may serve as a potential causal factor in shaping the intentions to use credit cards and amass credit card debt.

Additionally, empirical evidence solidifies the notion that there exists a noteworthy association between financial literacy and credit card debt, providing further support for the idea that financial literacy does indeed exert an influence on tangible financial behaviors (Brian, 2013).

### 2.7.2. Rational Addiction Theory

Rational Addiction Theory, initially proposed by economists Gary S. Becker and Kevin M. Murphy in 1988, offers a unique perspective on the consumption behavior of addictive goods and behaviors [13]. The theory suggests that individuals make rational decisions regarding addictive behaviors, such as excessive credit card use (Becker & Murphy, 1988). According to this theory, consumers may intentionally accumulate credit card debt when they anticipate future benefits, such as rewards points or discounts, will outweigh the costs of borrowing. The Theory assumes that individuals are rational actors who make choices based on maximizing their utility over time. In the context of credit card debt, this implies that consumers weigh the benefits and costs associated with borrowing on their credit cards, considering both immediate and future consequences.

Rational Addiction Theory implies that credit card debt accumulation may follow a dynamic pattern. As individuals become habituated to using credit cards for certain expenditures, they may gradually increase their debt load, especially if they perceive immediate benefits from doing so. The theory suggests that consumers may strategically accumulate credit card debt if they anticipate future benefits outweighing the costs. For example, if they expect to earn significant rewards points or cashback bonuses, they may be more inclined to use their credit cards for transactions.

## 3. METHODOLOGY

To achieve the objectives of the study, OLS regression analysis was used to estimate data gathered on credit card debt, GDP, and personal consumption. Ordinary Least Squares (OLS) regression analysis is an appropriate methodology for investigating the intricate relationship between credit card debt, GDP, and personal consumption in the United States. Its suitability lies in its robust handling of Quantitative Analysis, Statistical Validity, Causality Determination, Interpretability, and Time Series Data Analysis. This method aligns perfectly with the study's primary objective: to gauge the influence of credit card debt on personal consumption. OLS regression's strength is in precisely measuring and evaluating the strength and direction of these connections, shedding light on potential causal relationships without assuming causation based on correlation alone. The study's use of time series data spanning from 1999 to 2022 finds an ideal ally in OLS regression. This method excels in examining longitudinal data, enabling a comprehensive analysis of the shifts in credit card debt, GDP, and personal consumption over time. By doing so, it offers invaluable insights into the evolution and interconnectedness of these variables throughout the years, aiding in the understanding of their dynamics. OLS regression emerges as an invaluable tool in this study, empowering a thorough investigation into the complex associations between credit card debt, GDP, and personal consumption, thereby substantiating the study's objectives and contributing to a deeper comprehension of these critical economic variables.

The data were sourced secondarily from database of Federal Reserve of New York. Time series data which span between 1999 and 2022 was used in the research. Descriptive analysis (line graph) is also used to analyze the trend and relationship between the three variables (credit card debt, GDP, and personal consumption).

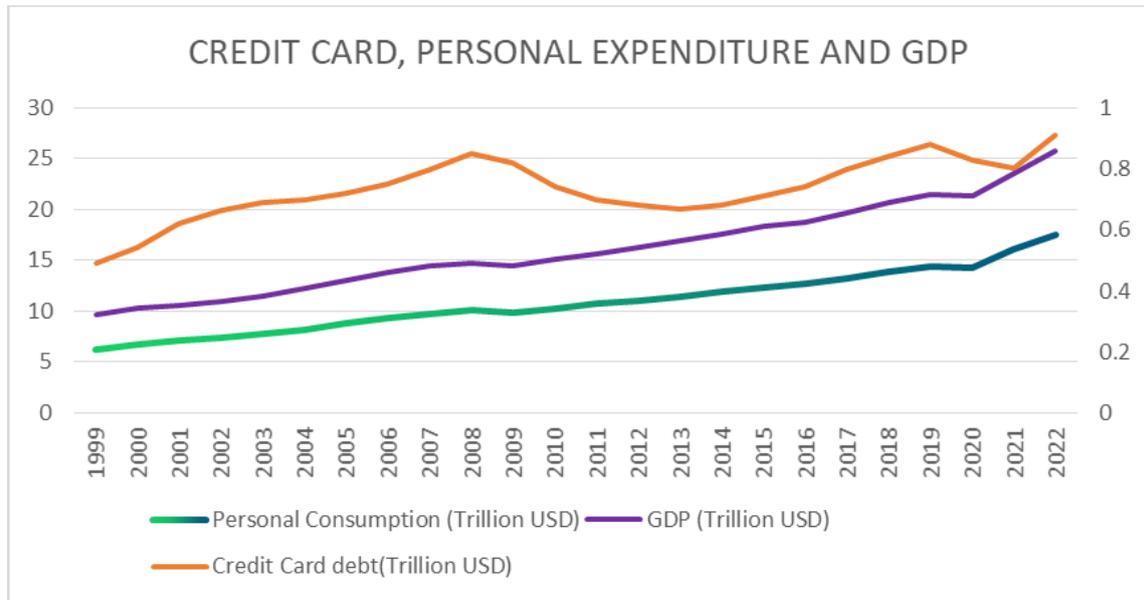

Figure 1.  Credit Card, Personal Consumption and GDP

Over the years, value of personal expenditure or consumption in US experience a steady increase year on year except in 2020 where total value of personal consumption experiences a sharp decline, this is attributed to the Covid-19 crisis. In 2022, total value of personal consumption in the economy is about $17 trillion. The growth in the economy characterized by GDP also has the same trend as total consumption but was greater than value of consumption over the years. GDP also experience decline to about $21 trillion in 2020 from about $22 trillion in 2019. The value of economic output start to increase from 2021 to about $25 trillion in 2022. The trend in credit card debt, unlike the trend in GDP and individual consumption experiences fluctuation over the years. There was consistent increase in value of credit card debt from almost $0.5 trillion in 1999 to about $0.9 trillion in 2008. Value of credit card debt was on decline from 2009 to 2013. In 2014, the value of credit card debt starts to increase and reached peak of above $0.9 trillion in 2019. Covid-19 crisis also affected total credit card debt as there was decline in the value before it increases to close to $1 trillion in 2022.

Table 1.  Regression Result (Impact of Credit Card Debt on Individual Consumption).

| Variables | Coefficients | T-Statistics | Prob. Value | |
|---|---|---|---|---|
| Credit Card Debt | 22.39166 | 5.390431 | 0.0000 | |
| C | -5.580041 | -1.812043 | 0.0000 | |
| Dependent Variable | Personal Consumption | | Sample: | 1999-2022 |
| R- squared | 0.569107 | | Adjusted R-squared | 0.549521 |
| F-statistic | 29.05675 | | Prob(F-statistic) | 0.000021 |
| *Indicates significance at 5% significance level | | | | |

Source: Author's Computation (2023) Using E-views 12.0 Software

The results of the regression analysis that shows the impact and relationship between credit card debt and individuals (personal consumption) in US shows that credit card debt have positive impact and relationship with personal consumption in US. The coefficient of the independent variable (22.39166) implies that a trillion dollar increase in value of credit card debt will result in $22.39 trillion increase in the value of personal consumption. The impact is significant as the probability value (0.000) of the independent variable (credit card debt) is less than 5% significance level. This implies that credit card debt has positive and significant impact on individual consumption in USA.

Table 2. Regression Result (Impact of Credit Card Debt on GDP).

| Variables | Coefficients | T-Statistics | Prob. Value | |
|---|---|---|---|---|
| Credit Card Debt | 32.52405 | 5.288504 | *0.0000 | |
| C | -7.785900 | -1.707776 | 0.1018 | |
| Dependent Variable | GDP | | Sample: | 1999-2022 |
| R- squared | 0.559721 | | Adjusted R-squared | 0.539708 |
| F-statistic | 27.96827 | | Prob(F-statistic) | 0.000026 |
| *Indicates significance at 5% significance level | | | | |

Source: Author's Computation (2023) Using E-views 12.0 Software

In the results of the regression analysis, credit card debt has positive impact on economic output (GDP) of US. A trillion dollar increase in value of credit card debt will result in $32.5 trillion increase in the value of GDP. The impact is significant as the probability value (0.000) of the independent variable (credit card debt) is less than 5% significance level. This implies that credit card debt has positive and significant impact on the US economy.

## 4. LIMITATIONS

While the regression analysis found significant relationships between credit card debt and both personal consumption and GDP, it's important to acknowledge that correlation does not confirm causation. External factors, such as changes in financial policies, socio-economic trends, and technological advancements, may influence credit card usage, and these were not extensively explored. More comprehensive analysis encompassing additional determinants, such as interest rates, employment trends, and technological advancements, will provide a deeper understanding of credit card debt dynamics. Furthermore, expanding the analysis to compare credit card debt dynamics across various countries or regions would offer broader insights into its economic implications. This comparative approach could reveal diverse trends and aid in understanding the varied impacts of credit card debt on different economies.

## 5. CONCLUSIONS

This paper has provided a comprehensive overview of the evolution of credit cards in the United States, their causes and consequences, and their impact on both individuals and the economy. It has examined the historical development of credit cards, the factors contributing to credit card debt, the consequences of credit card debt on individuals, and the broader economic implications. Additionally, it has explored strategies to prevent and manage credit card debt and presented empirical evidence to support the findings.

The historical evolution of credit cards in the United States highlights their transformation from specialized merchant cards to general-purpose credit cards that have become an integral part of everyday financial transactions. The legal changes, such as the Marquette decision, played a pivotal role in shaping the credit card industry's growth and accessibility to consumers.

Credit card debt has emerged as a significant financial challenge for many Americans, driven by economic factors, high healthcare costs, consumerism, financial illiteracy, and easy access to credit. The consequences of credit card debt on individuals are far-reaching, affecting their financial well-being, credit scores, savings, and even their physical and mental health. It can also strain personal relationships and limit opportunities for financial growth.

On a broader scale, credit cards have had a substantial impact on the U.S. economy. They stimulate consumer spending, drive e-commerce growth, and generate substantial revenue for financial institutions. However, they can also contribute to economic instability if not managed responsibly, as seen during the 2008 financial crisis.

To prevent and manage credit card debt, individuals can adopt various strategies, including financial education, budgeting, emergency savings, responsible credit card use, and seeking professional counselling. These strategies can empower individuals to make informed financial decisions and avoid falling into debt traps.

Empirical studies have supported the relationship between credit card debt and various factors, such as financial literacy, consumer behavior, and currency demand. The Theory of Planned Behavior and Rational Addiction Theory offer valuable frameworks for understanding and predicting credit card usage and debt accumulation.

Furthermore, regression analysis using historical data has shown that credit card debt has a significant positive impact on both personal consumption and GDP in the United States. This implies that as credit card debt increases, both individual spending and the overall economy tend to grow. However, it is essential to manage this debt responsibly to prevent negative consequences.

In conclusion, credit cards have become an integral part of the U.S. financial landscape, offering convenience and benefits to consumers. However, the responsible use of credit cards is crucial to avoid the pitfalls of excessive debt. By understanding the causes and consequences of credit card debt, individuals and policymakers can make informed decisions to promote financial well-being and economic stability.

The paper highlights the profound impact of credit card debt on both individuals and the broader U.S. economy. It offers a set of comprehensive recommendations to tackle the challenges associated with credit card debt while maximizing the benefits of credit card usage. These recommendations encompass various aspects of financial management and policy.

The recommendations include enhancing financial education and literacy programs, emphasizing responsible credit card use and transparency, promoting budgeting and financial planning, and encouraging the creation of emergency savings funds. They also advocate for lower-interest rate credit cards, regular monitoring of credit card statements, and access to professional financial counseling. Furthermore, the paper suggests policy reforms to enhance consumer protection, ongoing research to understand credit card debt dynamics, and efforts to promote financial inclusion. Public awareness campaigns and collaboration among stakeholders are also seen as essential components of addressing credit card debt challenges. Ultimately, these recommendations aim to transform credit cards into tools that empower individuals financially and contribute to economic stability, rather than sources of financial stress.

Future research should include an extended longitudinal study that goes beyond 2022 to capture recent events and economic changes. This research should delve into the post-pandemic landscape, analyzing how emerging financial trends, economic recovery, and evolving consumer behaviors impact credit card debt, GDP, and personal consumption. This study would provide critical insights into the lasting effects of economic crises on credit card dynamics and offer a real-time understanding of the post-pandemic financial landscape.

## AUTHORS

**Mayowa Jesse Akinwande** is a results-driven professional with a strong background in operations, IT, and data analysis. He holds a Bachelor of Science (BSc.) degree in Physics and Education, showcasing his commitment to both technical and educational pursuits. Currently, Mayowa is pursuing a Master of Science (MSc.) degree in Computer Science and Quantitative Methods, furthering his expertise in the ever-evolving field of technology and data analytics. Mayowa aims to leverage his data-driven skills and insights to drive positive change on a global scale. He is passionate about improving 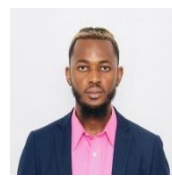 financial inclusion, operational excellence and using data-driven methodologies to deliver impactful projects worldwide.

**Alexander Toluwani Lopez** is a seasoned data professional with a diverse academic background and over 5 years of experience. He holds a Bachelor of Science in Electrical & Electronics Engineering from the University of Lagos, Nigeria, and is pursuing a master's in computer science and quantitative Methods with a focus on Predictive Analytics at Austin Peay State University, USA. His expertise in data analysis and software development, coupled with a strong passion for data analytics and machine learning, makes him an asset in the field of data science.

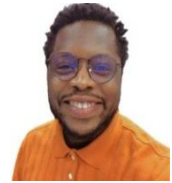

**Tobi Yusuf** has a Bachelor's degree, Geography and Earth Sciences/Geosciences, He is pursuing a master's degree in computer science and quantitative methods, with a concentration predictive analytics, Tobi is a skilled Quantitative Analyst with expertise in Manufacturing, Six Sigma and Data science.

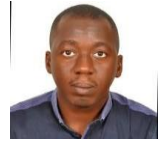

**Austine O. Unuriode** has a bachelor's degree in mathematics (Nigeria). He is currently pursuing a master's degree in computer science and quantitative methods, with a concentration in database management and analysis (USA). Austine has over 5 years of working experience as a data and business analyst. He has a keen interest in data analytics and data engineering, particularly in cloud computing.

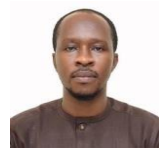

**Babatunde Y. Yusuf**, driven by an enduring passion for data analysis, graduated with a bachelor's degree in computer science from the University of Ibadan. Presently pursuing a Master of Science (MSc.) degree in Computer Science and Quantitative Methods at Austin Peay State University, whose focus lies in Predictive Analytics, covering skills like Python, R, and Machine Learning. Outside of this realm, he's been a loving husband since 2019 and takes pride in parenthood with his son, illustrating his commitment to both data and family.

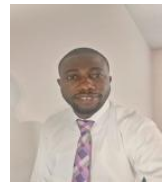

**Toyyibat T. Yussuph** holds a bachelor's degree in management and accounting (Nigeria) and a master's degree in management information systems (USA). Her extensive experience in product development, data strategy, and financial and data analysis has enabled her to transform and automate several critical insights into fraud trends and risk mitigation strategies as a Product Development Manager with American Express. She has a passion for innovation and leveraging big data to solve Credit and Fraud risks within the Financial Service industry.

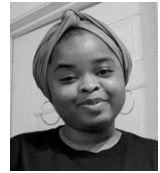

**Stanley Okoro** is a highly skilled and innovative ICT Engineer with expertise in 5G, Cloud computing, and Generative AI. He has a bachelor's degree in electrical and Electronics Engineering, an MBA in Global Business Management, and is currently completing his second master's degree in computer science and quantitative Methods at Austin Peay State University, Clarksville Tennessee, USA. He has a deep passion for cutting-edge technology and specializes in bridging business objectives with technical solutions.

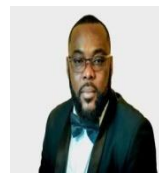